\documentclass[12pt]{iopart}
\usepackage{latexsym}
\usepackage{graphics}
\usepackage{graphicx}
\usepackage{amssymb}
\usepackage[latin1]{inputenc}

\begin{document}

\title{Superconducting Microwave Cavity Made of Bulk MgB$_2$}

\author{G. Giunchi}
\address{EDISON SpA R \& D, Foro Buonaparte 31, I-20121 Milano
(Italy)}

\author{A. Agliolo Gallitto, G. Bonsignore, M. Bonura, M. Li Vigni}
\address{CNISM and Dipartimento di Scienze Fisiche e
Astronomiche, Università di Palermo, Via Archirafi 36, I-90123
Palermo (Italy)}

\begin{abstract}
We report the successful manufacture and characterization of a
microwave resonant cylindrical cavity made of bulk MgB$_2$
superconductor ($T_c \approx 38.5~$K), which has been produced by
the Reactive Liquid Mg Infiltration technique. The quality factor
of the cavity for the TE$_{011}$ mode, resonating at 9.79 GHz, has
been measured as a function of the temperature. At $T=4.2$~K, the
unloaded quality factor is $\approx 2.2 \times 10^5$; it remains
of the order of $\times 10^5$ up to $T \sim 30$~K. We discuss the
potential performance improvements of microwave cavities built
from bulk MgB$_2$ materials produced by reactive liquid Mg
infiltration.
\end{abstract}



\maketitle

The very low surface resistance of superconducting materials makes
them particularly suitable for designing high-performance
microwave (mw) devices, with considerably reduced sizes. The
advent of high-temperature superconductors (HTS) further improved
the expectancy for such applications, offering a potential
reduction of the cryogenic-refrigeration limit, with respect to
low-temperature superconductors. A comprehensive review on the mw
device applications of HTS was given by Lancaster \cite{lanc}.
Among the various devices, the superconducting resonant cavity is
one of the most important applications in the systems requiring
high selectivity in the signal frequency, such as filters for
communication systems \cite{pand}, particle accelerators
\cite{padam,collings}, equipments for material characterization at
mw frequencies \cite{lanc,zhai}.

Nowadays, one of the commercial applications of HTS electronic
devices are planar-microstrip filters for transmission line, based
on YBa$_2$Cu$_3$O$_7$ thin or thick films \cite{lanc,hein}, whose
manufacturing process allows having an high degree of device
miniaturization. Nevertheless, the need of high performance, in
many cases, overcomes the drawback of the device sizes, as, e.g.,
in the satellite-transmission systems, radars, particle
accelerators, and demands for cavities with the highest quality
factors. Since the discovery of HTS, several attempts have been
done to manufacture mw cavities made of these materials in bulk
form \cite{pand,zaho,lanc92}; however, limitations in the
performance were encountered. Firstly, because of the small
coherence length of HTS, grain boundaries in these materials are
weakly coupled giving rise to reduction of the critical current
and/or nonlinear effects \cite{golo}, which worsen the device
performance; furthermore, the process necessary to obtain bulk HTS
in a performing textured form is very elaborated. For these
reasons, in several applications, such as particle accelerators
and equipments for mw characterization of materials, most of the
superconducting cavities are still manufactured by Nb, requiring
liquid helium as refrigerator.

Since the discovery of superconductivity at 39~K in
MgB$_2$~\cite{naga}, several authors have indicated this material
as promising for technological
applications~\cite{collings,bugo,HeinProc}. Indeed, it has been
shown that bulk MgB$_2$, contrary to oxide HTS, can be used in the
polycrystalline form without a significant degradation of its
critical current \cite{bugo,HeinProc,larbalestier}. This property
has been ascribed to the large coherence length, which makes the
material less susceptible to structural defects like grain
boundaries. Actually, it has been shown that in MgB$_2$ only a
small amount of grain boundaries act as weak
links~\cite{Samanta,Rowell,Khare,agli2}. Furthermore, MgB$_2$ can
be processed very easily as high-density bulk material
\cite{giun03}, showing very high mechanical strength. Due to these
amazing properties, MgB$_2$ has been recommended for manufacturing
mw cavities~\cite{collings,taji}, and investigation is carried out
to test the potential of different MgB$_2$ materials for this
purpose. However, papers discussing the realization and/or
characterization of mw cavities made of MgB$_2$ have not yet been
reported.

Recently, we have investigated the mw response of MgB$_2$ samples
prepared by different methods, in the linear and nonlinear
regimes~\cite{agli2,agli}. Our results have shown that the
residual surface resistance strongly depends on the preparation
technique and the purity and/or morphology of the components used
in the synthesis process. In particular, the investigation of
small plate-like samples of MgB$_2$ prepared by Reactive Liquid Mg
Infiltration (RLI) process, has highlighted a weak nonlinear
response, as well as relatively small values of the residual
surface resistance. Furthermore, bulk samples produced by RLI
maintain the surface staining unchanged for years, without
controlled-atmosphere protection. This worthwhile property is most
likely related to the high density, and consequently high grain
connectivity, achieved with the RLI process, as well as to the
small and controlled amount of impurity phases
\cite{giun-physicaC}. On the contrary, samples prepared by other
techniques, though exhibiting lower values of the residual surface
resistance, need to be kept in protected atmosphere to avoid their
degradation. These interesting results have driven us to build a
mw resonant cavity using MgB$_2$ produced by the RLI process. In
this work, we discuss the properties of the first mw resonant
cavity made of bulk MgB$_2$.

As the first attempt to apply the MgB$_2$ to the cavity-filter
technology, we have manufactured a simple cylindrical cavity and
have investigated its microwave response in a wide range of
temperatures. All the parts of the cavity, cylinder and lids, are
made of bulk MgB$_2$ with $T_c \approx 38.5$~K and density
$\approx 2.33~\mathrm{g/cm^3}$. The MgB$_2$ material has been
produced by the RLI process \cite{giun03,giun-cryo06}, which
consists in the reaction of B powder and pure liquid Mg inside a
sealed stainless steel container. In particular, the present
cylindrical cavity (inner diameter 40 mm, outer diameter 48 mm,
height 42.5 mm) was cut by electroerosion from a thicker bulk
MgB$_2$ cylinder prepared as described in Sec.~4.3 of
Ref.~\cite{giun-cryo06}, internally polished up to a surface
roughness of about 300~nm. A photograph of the parts, cylinder and
lids, composing the superconducting cavity is shown in Fig.~1.

\begin{figure}[htbp]
\centering
\includegraphics[width=0.6\textwidth]{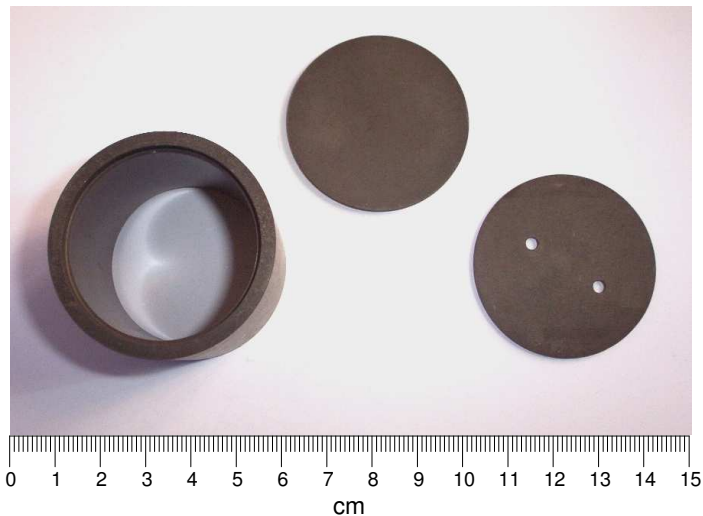}
\caption{Photograph of the cylinder and the lids composing the
bulk MgB$_2$ cavity. The holes in one of the lids are used to
insert the coupling loops.}
\label{cavity}
\end{figure}

As it is well known, resonant cylindrical cavities support both TE
and TM modes. In the TE$_{01n}$ modes, the wall currents are
purely circumferential and no currents flow between the lids and
the cylinder, requiring no electrical contact between them; for
this reason, the TE$_{01n}$ are the most extensively used modes.
The TE$_{01n}$ modes are degenerate in frequency with the
TM$_{11n}$ modes and this should be avoided to have a well defined
field configuration. In order to remove the degeneracy, we have
incorporated a ``mode trap" in the form of circular grooves (1 mm
thick, 2 mm wide) on the inside of the cylinder at the outer
edges. This shifts the resonant frequency of the TM$_{11n}$ modes
downwards, leaving the TE$_{01n}$ modes nearly unperturbed.

Two small loop antennas, inserted into the cavity through one of
the lids, couple the cavity with the excitation and detection
lines. The loop antenna was constructed on the end of the lines,
soldering the central conductor to the outer shielding of the
semirigid cables.

The ratio between the energy stored in the cavity and the energy
dissipated determines the quality factor, $Q$, of the resonant
cavity. When the cavity is coupled to an external circuit, besides
the power losses associated with the conduction currents in the
cavity walls, additional losses out of the coupling ports occur.
The overall or loaded $Q$ (denoted by $Q^L$) can be defined by
\begin{equation}
\frac{1}{Q^L}=\frac{1}{Q^U}+\frac{1}{Q^R}\,,
\end{equation}
where $Q^U$ is the so-called unloaded $Q$, determined by the
cavity-wall losses and $Q^R$ is due to the effective losses
through the external coupling network.

The resonant frequency and unloaded quality factor of a
cylindrical cavity resonating in the TE$_{01n}$ mode are given by
\cite{lanc,lanc92}
\begin{equation}
f_{01n} = \frac{1}{2 \pi \sqrt{\epsilon \mu}}
\sqrt{\left(\frac{n\pi}{d}\right)^2+\left(\frac{z_{01}}{a}\right)^2}\,,
\end{equation}

\begin{equation}
Q_{01n}^U = \frac{1}{R_s}\,\sqrt{\frac{\mu}{\epsilon}}
\frac{\left[(z_{01}d)^2 + (n \pi a)^2\right]^{3/2}}{2z_{01}^2 d^3
+ 4 n^2\pi ^2 a^3}\,,
\end{equation}
where $\mu$ and $\epsilon$ are the permeability and dielectric
constant of the medium filling the cavity; $a$ and $d$ are the
radius and length of the cavity; $R_s$ is the surface resistance
of the material from which the cavity is built; $z_{01} = 3.83170$
is the first zero of the derivative of the zero-order Bessel function.\\
$Q^U$ can be determined by taking into account the coupling
coefficients, $\beta_1$ and $\beta_2$ for both the coupling lines;
these coefficients can be calculated by directly measuring the
reflected power at each line, as described in Ref.~\cite{lanc},
Chap. IV. Thus, $Q^U$ can be calculated as
\begin{equation}
Q^U = (1+\beta_1+\beta_2)Q^L\,.
\end{equation}

The frequency response of the cavity has been measured in the
range of frequencies 8~$\div$~13~GHz by an \textit{hp}-8719D
Network Analyzer. Transmission by two probes has been successfully
used for measuring the loaded quality factor in a wide range of
temperatures. Among the various modes detected, two of them have
shown the highest quality factors; at room temperature and with
the cavity filled by helium gas, the resonant frequencies of these
modes are 9.79~GHz and 11.54~GHz; according to Eq.~(2), they
correspond to the TE$_{011}$ and TE$_{012}$ modes. The coupling
coefficients $\beta_1$ and $\beta_2$ have been measured as a
function of the temperature; they result $\approx 0.2$ at $T =
4.2$~K and reduce to $\approx 0.05$ when the superconductor goes
to the normal state. At $T=4.2$~K (without liquid helium inside
the cavity) the unloaded quality factors, determined by using
Eq.~(4), are $Q_{011}^U\approx 220000$ and $Q_{012}^U\approx
190000$; both decrease by a factor $\approx 20$ when the material
goes to the normal state.

\begin{figure}[htbp]
\centering
\includegraphics[width=0.65\textwidth]{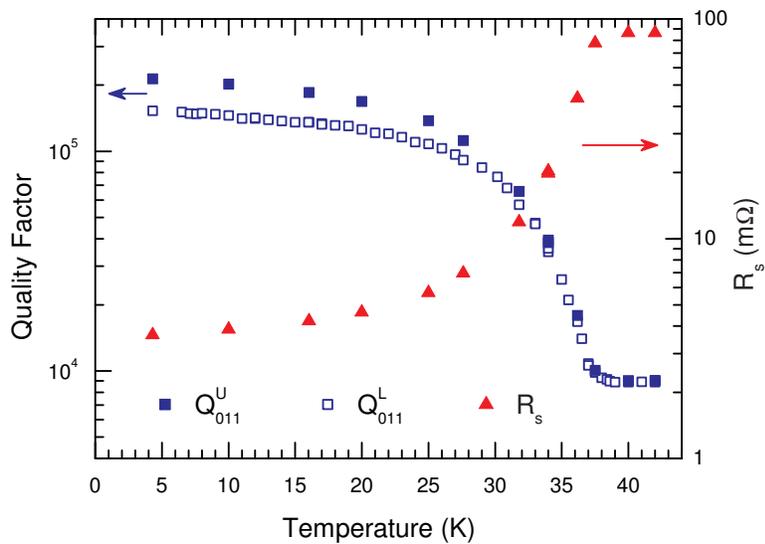}
\caption{Temperature dependence of the loaded and unloaded quality
factor, $Q_{011}^L$ and $Q_{011}^U$, (left scale); mw surface
resistance $R_s$ deduced from $Q_{011}^U$ (right scale).}
\label{quality}
\end{figure}

Fig.~\ref{quality} shows the temperature dependence of the
measured (loaded) and the calculated (unloaded) $Q$ values for the
TE$_{011}$ mode, $Q_{011}^L$ and $Q_{011}^U$; in the same plot
(right scale) it is shown the mw surface resistance deduced from
$Q_{011}^U$ using Eq.~(3). As one can note, the quality factor
maintains values of the order of $10^5$ up to $T \approx 30$~K and
reduces by a factor $\approx 20$ at $T=T_c$.

The results of Fig.~\ref{quality} have been obtained at low input
power level ($\approx -15$~dBm). In order to reveal possible
nonlinear effects, we have investigated the TE$_{011}$ resonant
curve at higher power levels. In this case, to avoid possible
heating effects, the measurements have been performed with the
cavity immersed into liquid helium. At input power level of
$\approx 15$~dBm, we have observed that the quality factor reduces
by about 10\%, indicating that, at these input power levels,
nonlinear effects on the surface resistance are weak.

Our results show that MgB$_2$ produced by RLI is a very promising
material for building mw resonant cavities. We have obtained a
quality factor higher than those reported in the literature for mw
cylindrical cavities manufactured by HTS, both bulk and films
\cite{pand,hein,zaho,lanc92}. Moreover, $Q$ takes on values of the
order of $10^5$ from $T=4.2$~K up to $T\approx 30$~K, temperature
easily reachable by modern closed-cycle cryocoolers.

We would remark that this is the first attempt to realize a
superconducting cavity made of bulk MgB$_2$; we expect that the
performance would improve if the cavity were manufactured with
material produced by liquid Mg infiltration in micrometric B
powder. The present cavity is made of MgB$_2$ material obtained
using crystalline B powder with grain mean size $\approx
100~\mu$m. On the other hand, previous studies have shown that the
grain size of the B powder, used in the RLI process, affects the
morphology \cite{giun-IEEE-06} and the superconducting
characteristics \cite{agli,giun-IEEE-06} of the material,
including the mw properties.

Investigation of the microwave response of bulk MgB$_2$ obtained
by the RLI method has been performed in the linear and nonlinear
regimes \cite{agli,agli2}. In the linear regime, we have measured
the temperature dependence of the mw surface impedance \cite{agli}
at 9.4~GHz; the results have shown that the sample obtained using
microcrystalline B powder ($\approx 1~\mu$m in size) exhibits
smaller residual surface resistance ($< 0.5~\mathrm{m} \Omega$)
than those measured in samples prepared by crystalline B powder
with larger grain sizes \cite{agli}. Since the residual surface
resistance obtained from the $Q_{011}^U$ data is
$R_s(4.2~\mathrm{K})\approx 3.5~\mathrm{m}\Omega$, we infer that
by using microcrystalline B powder in the RLI process the $Q$
factor could increase by one order of magnitude.

In the nonlinear regime (input peak power $\sim 30$~dBm), we have
investigated the power radiated at the second-harmonic frequency
of the driving field \cite{agli2}. Since it has been widely shown
that the second-harmonic emission by superconductors at low
temperatures is due to nonlinear processes in weak links
\cite{golo,samoilova,lee}, these studies allow to check the
presence of weak links in the samples. Our results have shown that
MgB$_2$ samples produced by RLI exhibit very weak second-harmonic
emission at low temperatures. In particular, the sample obtained
using microcrystalline B powder does not show detectable
second-harmonic signal in a wide range of temperatures, from
$T=4.2$~K up to $T \approx 35$~K \cite{agli2}. So, we infer that
eventual nonlinear effects in the cavity response can be reduced
by using microcrystalline B powder.

Because of the shorter percolation length of the liquid Mg into
very fine B powder ($1~\mu$m in size), the production of massive
MgB$_2$ samples by RLI using microcrystalline B powder turns out
to be more elaborated. In this work, we have devoted the attention
to explore the potential of bulk MgB$_2$ materials prepared by RLI
for manufacturing mw resonant cavities; work is in progress to
improve the preparation process in order to manufacture large
specimens using microcrystalline B powder.

In summary, we have successfully built and characterized a mw
resonant cavity made of bulk MgB$_2$. We have measured the quality
factor of the cavity for the TE$_{011}$ mode as a function of the
temperature, from $T=4.2$~K up to $T \approx 45$~K. At $T=4.2$~K,
the unloaded quality factor is $Q_{011}^U \approx 2.2 \times
10^5$; it maintains values of the order of $10^5$ up to $T \sim
30$~K and reduces by a factor $\approx 20$ when the superconductor
goes to the normal state. To our knowledge, these $Q$ values are
larger than those obtained in HTS bulk cavities in the same
temperature range. The results show that the RLI process provides
a useful method for designing high-performance mw cavities, which
may have large scale application. We have also indicated a way to
further improve the MgB$_2$ mw cavity
technology.\\
\\
\noindent The authors acknowledge Yu. A. Nefyodov and A. F.
Shevchun for critical reading of the manuscript.

\end{document}